\documentclass[prl,twocolumn,aps,epsf]{revtex4}

\usepackage{amsmath}
\usepackage{epsfig}
\usepackage{color}
\usepackage{latexsym}
\usepackage{fancyhdr}
\usepackage{graphicx}
\usepackage{amsfonts}
\usepackage[latin1]{inputenc}
\usepackage{amssymb}
\usepackage[colorlinks=true, linkcolor=blue,
            citecolor=red, urlcolor=green]{hyperref}

\def\lesim{\ \hbox to 0 pt{\raise .6ex\hbox{$<$}} \lower .5ex\hbox{$\sim$}\ }
\def\gesim{\ \hbox to 0 pt{\raise .6ex\hbox{$>$}} \lower .5ex\hbox{$\sim$}\ }
\def\erfc{ \textrm{erfc} }
\def\erfcx{ \textrm{erfcx} }

\def\atan{ \textrm{atan} }
\def\acos{ \textrm{acos} }
\def\ext{ \textrm{ext} }

\def\d{ \textrm{d} }

\begin{document}

\title{Fully self-consistent ion-drag force calculations for dust in
collisional plasmas with external electric field}

\author{Leonardo Patacchini}

\author{Ian H. Hutchinson}

\affiliation{Plasma Science and Fusion Center and Department of
Nuclear Science and Engineering, Massachusetts Institute of
Technology, cambridge, MA 02139, USA}

%%%%%%%%%%%%%%%%%%%%%%%%%%%%%
%%%% Abstract %%%%%%%%%%%%%%%
%%%%%%%%%%%%%%%%%%%%%%%%%%%%%

\begin{abstract}

The ion-drag force on a spherical dust particle immersed in a flowing
plasma with external electric field is self-consistently calculated
using the Particle In Cell code SCEPTIC in the entire range of
charge-exchange collisionality. Our results, not based on questionable
approximations, extend prior analytic calculations valid only in a few
limiting regimes. Particular attention is given to the force
direction, shown never to be directed opposite to the flow except in
the continuum limit, where other forces are of much stronger
magnitude.

\end{abstract}

\maketitle

%%%%%%%%%%%%%%%%%%%%%%%%%%%%%
%%%% Introduction %%%%%%%%%%%
%%%%%%%%%%%%%%%%%%%%%%%%%%%%%

Understanding the behaviour of a single dust particle embedded in a
flowing plasma is a physical problem of practical relevance to the
field of dusty plasmas, with concrete applications in fusion-grade,
astrophysical, laboratory and technological
plasmas~\cite{DustInt}. The most important observable effect is
perhaps momentum transfer between the plasma and the dust grain ---the
ion-drag force~\cite{Khrapak1,Ivlev,HutchDrag}--- responsible for the
dust dynamics as well as the formation of static configurations. Force
calculations have very recently received renewed attention in the
context of weakly ionized plasmas, where effects of fundamental
physical importance transcending the field of dusty plasmas have been
proposed. It has for example been proposed that two positively charged
dust particles can attract each other by a process analogous to cooper
pairing in superconductors~\cite{Attraction}.

More surprisingly, it is found that in the same strongly collisional
limit the ion-drag on a single dust grain could be
negative~\cite{KhrapFluid1,HutchRep,KhrapRep}, implying that those
particles see the surrounding plasma as a
superfluid~\cite{Superfluid}. Several published calculations have been
done by treating the plasma as a linear dielectric medium, and are
therefore only valid for large shielding lengths $\lambda_s\gg
b_{90}$, where $b_{90}$ is the $90^o$ scattering parameter of the dust
particle. Only the weak and moderate~\cite{Ivlev} as well as strongly
collisional~\cite{KhrapFluid1} regimes have been investigated with
this approach, and the results of Ref.~\cite{Ivlev} are not directly
quantifiable because they depend on the dust floating charge $Q_f$,
not known \emph{a priori}. In the highly collisional limit, full
non-linear calculations have been performed under the assumption of
negligible ion diffusivity (mobility regime)~\cite{HutchRep,KhrapRep},
predicting that ion-drag reversal only occurs at strong
collisionality. Some Monte-carlo simulations~\cite{Schweigert} claim a
negative ion-drag occurs also in the weak collisionality regime.

The purpose of this letter is to calculate the ion-drag force on a
conducting spherical dust particle (radius $r_p$) in the entire range
of collisionality, by solving the full self-consistent non linear
problem; this bridges the gaps between all available partial
results. For this purpose we use the Particle In Cell (PIC) code
SCEPTIC~\cite{HutchDrag}, recently upgraded to account for constant
collision-frequency ($\nu_c$) charge-exchange
events~\cite{ScepticColl}. More complex collision models could be
implemented, but the present approach limits the number of parameters
while only negligibly affecting the physics~\cite{Rovagnati}. SCEPTIC
is a hybrid kinetic code where the collisionless electrons have the
Boltzmann density $n_e=n_{\infty}\exp(\phi)$ ($\phi=eV/T_e$ and
$n_{\infty}$ is the electron density at infinity) and the ions are
advanced according to Newton's equation between two collisions
($m_i\d{\bf v}/d t=-Ze{\bf \nabla}V$) up to convergence of the
simulation. The electrostatic potential is solved on a two-dimensional
spherical mesh centered on the dust particle and extending several
electron Debye lengths
$\lambda_{De}=\sqrt{\epsilon_0T_e/n_{\infty}e^2}$. The dust potential
can be prespecified or floating, i.e. self-consistently set to a value
such as to equate the ion current with the analytic collisionless
electron current $I_e=4\pi r_p^2 v_{te}/2\sqrt{\pi}$
($v_{te}=\sqrt{2T_e/m_e}$ is the electron thermal speed). At the outer
boundary we impose $\d\ln\phi/\d r=-r_p(1/(\lambda_sr)+1/r^2)$, a
condition valid at weak and moderate collision frequency as well as in
the continuum limit if $\lambda_s\gg r_p$.

%%%%%%%%%%%%%%%%%%%%%%%%%%%%%
%%%% Ion Distributions %%%%%%
%%%%%%%%%%%%%%%%%%%%%%%%%%%%%

%If the ion drift were driven by a background
%neutral flow ($v_n=v_d$), this distribution would be a drifting
%Maxwellian 
%\begin{equation}
%  f_M({\bf v})
%  =\frac{n_{\infty}}{(v_{ti}\sqrt{\pi})^3}\exp\left(-\left(\frac{{\bf v}-{\bf
%  v_d}}{v_{ti}}\right)^2\right),\label{fionDn}
%\end{equation}

The ions are reinjected at the outer boundary according to their
distribution function at infinity (``SCEPTIC1'' convention in
Ref.~\cite{ScepticColl}). We focus on the situation where the neutral
background is stationary and the ion flow driven by an external
electric field ${\bf E}_{\ext}={\bf v_d}m_i\nu_c/Ze$ (Z is the ion
charge, usually Z=+1), whose role is to compensate the ion-neutral
friction at infinity. In this case, provided $\nu_{ii}\ll\nu_c$
($\nu_{ii}$ is the ion-ion Coulomb collision frequency), the ion
distribution at infinity is given by~\cite{Fahr}:
%\begin{equation}
%  f_i^{\infty}({\bf v})=\frac{n_{\infty}}{(v_{ti}\sqrt{\pi})^3}
%  \int_{t=0}^{\infty}\exp\left(-\left(\frac{{\bf v}-{\bf v_d}\cdot t
%  }{v_{ti}}\right)^2\right)\exp(-t)\d t,
%\end{equation}
\begin{equation}
  f_i^{\infty}({\bf v})=\frac{1}{(v_{ti}\sqrt{\pi})^2}\frac{1}{2v_d}
  \exp(-\frac{{\bf v}^2}{v_{ti}^2})\erfcx
  \left(\frac{v_{ti}}{2v_d}-\frac{v_z}{v_{ti}}\right)\label{fionSn},
\end{equation}
where $\erfcx(x)=\exp(x^2)\erfc(x)$. $f_i^{\infty}$ does not depend on
$\nu_c$, and tends to a drifting Maxwellian $f_{M}$ with temperature
$T_i$ and thermal speed $v_{ti}=\sqrt{2T_i/m_i}$ as
$v_d/v_{ti}\rightarrow 0$. Here $T_i$ refers to the temperature of the
neutral background; the effective ion temperature is
$T_{i,z}^{\infty}=T_i+m_iv_d^2\geq T_i$ and
$T_{i,\perp}^{\infty}=T_i$.

%Provided $\nu_{ii}\ll\nu_c$, where
%$\nu_{ii}$ is the ion-ion Coulomb collision frequency, we reinject the
%ions according to Eq.~(\ref{fionSn}) even in the \emph{collisionless}
%limit $\nu_c\ll\sqrt{ZT_e/m_i}/r_p$.

%%%%%%%%%%%%%%%%%%%%%%%%%%%%%%

The ion-drag ${\bf F_i}$ is the sum of the electrostatic force on the
dust particle surface ${\bf F_E^p}$ arising from the interaction of
its usually negative charge with the flow-induced anisotropy of the
plasma (calculated by integration of the Maxwell stress tensor at the
particle surface), the momentum collected by direct ion impact ${\bf
F_{im}^p}$, and the electron-pressure force ${\bf F_e^p}=0$ (averaging
to zero on a conducting body~\cite{HutchDrag}): ${\bf F_i}={\bf
F_{im}^p}+{\bf F_E^p}$. Momentum conservation implies that in steady
state, the ion-drag be also equal to the rate of momentum flux across
any control surface surrounding the dust particle, in particular the
outer boundary of the computational domain~\cite{HutchDrag}. In a
collisionless plasma: ${\bf F_i}={\bf F_{im}^o}+{\bf F_E^o}+{\bf
F_e^o}$ (${\bf F_{im}^o}$: net ion momentum flux into the
computational domain; ${\bf F_E^o}$ and ${\bf F_e^o}$: integrals of
the electrostatic stress and electron pressure on the boundary). Of
course because the SCEPTIC domain is not infinite, the electrostatic
stress and electron pressures at the outer boundary are not
negligible. In the presence of ion-neutral collisions, one must also
consider ion friction with the neutrals and the momentum provided by
the external electric field inside the control volume:
\begin{equation}
  {\bf F_n^o}=m_i\nu_c\int_{\textrm{Domain}}n_i({\bf x})({\bf
v_d}-{\bf v}({\bf x}))\d^3{\bf x},\label{Fno}
\end{equation}
where ${\bf v}({\bf x})$ is the ion fluid (local average)
velocity. $F_n^o$ can be either positive or negative; the integral in
Eq.~(\ref{Fno}) is however convergent and $F_n^o$ tends to a limit as
the domain size is increased. We will refer to ${\bf F_{im}}$, ${\bf
F_E}$, ${\bf F_e}$, ${\bf F_n}$ respectively as the ``Ion'',
``E-field'', ``Electrons'' and ``Collisions'' forces.

%%%%%%%%%%%%%%%%%%%%%%%%%%%%%
%%%% Schweig Comp %%%%%%%%%%%
%%%%%%%%%%%%%%%%%%%%%%%%%%%%%

\begin{figure}[!htp]
  \vspace{-0.4cm}
  \includegraphics[width=8.5cm]{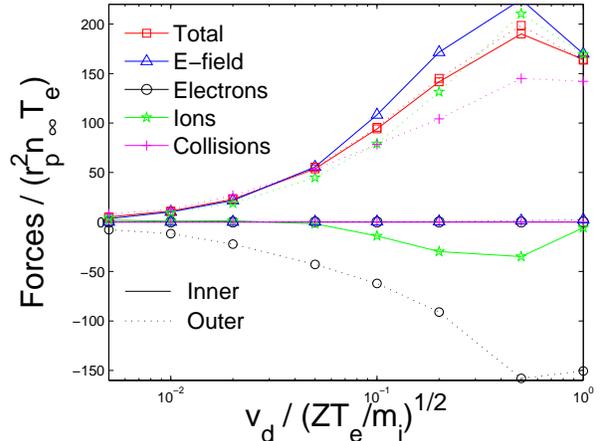}\\
  \vspace{-0.5cm}
  \caption{\small Ion-drag force as a function of drift velocity
    computed by SCEPTIC for $T_i=4.33\cdot 10^{-3}ZT_e$,
    $\lambda_{De}=64.8r_p$, $\phi_p=-1.49$ (Here imposed rather than
    floating), and $\nu_c=6.27\cdot 10^{-3}\sqrt{ZT_e/m_i}/r_p$;
    corresponding to the curve $\lambda_s=20\mu m$ and $P=150Pa$ in
    Fig.~(2) from Ref.~\cite{Schweigert}. ``Inner'' (solid lines) and
    ``Outer'' (dotted lines) refer to the forces calculated at the
    dust surface and outer boundary of the computational domain; of
    course the total force (squares) does not depend on where it is
    evaluated.\label{SchwRev}}
  \vspace{-0.2cm}
\end{figure}

We begin by comparing our code with the results of Schweigert and
coauthors~\cite{Schweigert}, who propose a weak collisionality regime
($\lambda_{mfp}=\sqrt{8T_i/\pi m_i}/\nu_c\gg \lambda_s$) for which
they find a negative ion-drag. The dimensional parameters
corresponding to their Fig.~(2) are $r_p=4.7\mu m$,
$\lambda_s=20-100\mu m$, $Q_p=3.6\cdot 10^4e$ (dust charge),
$T_e=6eV$, $T_i=0.026eV$, $P=75-150Pa$ (neutral pressure),
$\sigma_c=3.53\cdot 10^{-15}cm^2$ (constant collision cross-section),
$n_{\infty}=3.6\cdot 10^9cm^{-3}$. Helium ions are used, but this
information is irrelevant because their work is based on a direct
orbit-integration approach, where the dust charge as well as the
potential distribution are prespecified. If we use the usual
linearized formula for the sphere capacitance:
$C=4\pi\epsilon_0r_p(1+r_p/\lambda_s)$, for simplicity set
$\lambda_s^{-2}=\lambda_D^{-2}=\lambda_{De}^{-2}+\lambda_{Di}^{-2}$,
and convert constant cross-section into constant collision-frequency
according to $\nu_c=\sigma_c\sqrt{8T_i/\pi m_i}P/T_i$, we find the
corresponding parameters to use for our SCEPTIC
simulations. Fig.~(\ref{SchwRev}) shows the total ion-drag as well as
its different components for the case corresponding to the curve
$\lambda_s=20\mu m$ and $P=150Pa$ in Fig.~(2) of Schweigert's
publication~\cite{Schweigert}. The ion-drag calculated by SCEPTIC at
the dust surface and the outer boundary of the computational domain
agree (their balance is of course different), which gives us strong
confidence that our code performs properly and the runs are well
converged.
More important the total ion-drag is positive, and does not agree with
Schweigert's according to which it can reverse for low drift
velocities.

\begin{figure}[!htp]
  \vspace{-0.4cm}
    \includegraphics[width=8.5cm]{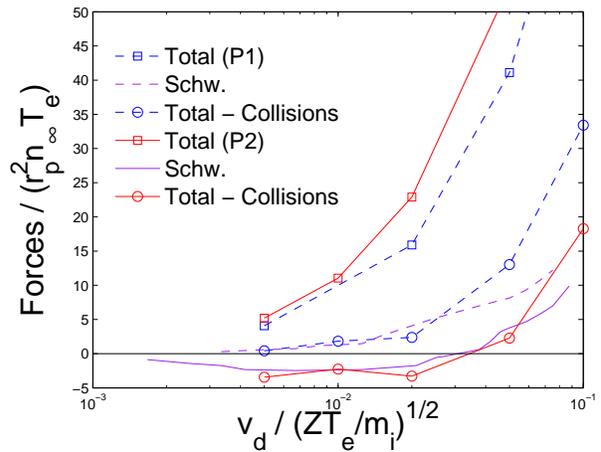}\\
    \vspace{-0.5cm}
    \caption{\small Ion-drags from Fig.~(2) in Ref.~\cite{Schweigert}
      at $\lambda_s=20\mu m$ (``Schw.''), ion-drags computed by
      SCEPTIC at the outer boundary of the computational domain
      (``Total''), and ion-drags computed by SCEPTIC without
      accounting for $F_n^o$ (``Total$-$Collisions''). $P_1$: $P=75Pa$
      i.e.  $\nu_c=3.14\cdot 10^{-3}\sqrt{ZT_e/m_i}/r_p$; $P_2$:
      $P=150Pa$ i.e.  $\nu_c=6.27\cdot 10^{-3}\sqrt{ZT_e/m_i}/r_p$
      ($P_2$ curves correspond to the parameters of
      Fig.~(\ref{SchwRev})). The \emph{correct} ion-drag is is always
      \emph{positive} (lines with square-markers).\label{SchwComp}}
    \vspace{-0.2cm}
\end{figure}

Our hypothesis to explain Schweigert's erroneous conclusion is as
follows. Fig.~(\ref{SchwComp}) shows the two ion-drag curves from
Fig.~(2) of Ref.~\cite{Schweigert} with $\lambda_s=20\mu m$ (lines),
as well as the ion-drag computed by SCEPTIC at the outer boundary of
the domain without accounting for $F_n^o$ (lines with circle-markers),
and the \emph{correct} ion-drag (lines with square-markers). The
strong similarity between Schweigert's results and SCEPTIC-computed
drags not accounting for $F_n^o$ suggests that the far-from-negligible
$F_n^o$ has been neglected in Ref.~\cite{Schweigert}.

%%%%%%%%%%%%%%%%%%%%%%%%%%%%%
%%%% Khrap Comp 2 %%%%%%%%%%%
%%%%%%%%%%%%%%%%%%%%%%%%%%%%%

The only regime where we have observed a negative ion-drag is when
$\nu_c\gg \sqrt{ZT_e/m_i}/r_p$. In this highly collisional limit a
fluid treatment is appropriate, and analytic calculations assuming
negligible diffusivity (i.e. mobility-dominated physics) and
$\lambda_s\gg r_p$ have recently been
published~\cite{HutchRep,KhrapRep}. According to this model the ion
density is uniform on the ion fluid stream-lines, hence the ion
current (required for an analytic calculation of the floating
potential) and the electrostatic stress at the particle surface only
depend on the stream-lines' topology~\cite{HutchRep,KhrapRep}:
\begin{equation}
  I_i=\pi n_{\infty} T_e r_p\frac{|\phi_p|}{m_i\nu_c}
  \left[\frac{1}{\tilde{a}_0^2}+2+\tilde{a}_0^2\right]\label{KhrapCurr}
\end{equation}
\begin{eqnarray}
  F_E^p=\pi r_p^2 n_{\infty} T_e
  |\phi_p|\left[(1-\tilde{a}_0^{-4})\right.\nonumber\\
  -\left.2\tilde{a}_0^{-1}\int_0^{\acos(\tilde{a}_0^{-2})}
  \sqrt{\tilde{a}_0^2 +\tilde{a}_0^{-2}-
  2\cos\theta}\cos\theta\d\theta\right],\label{KhrapFE}
\end{eqnarray}
where $\tilde{a}_0=a_0/r_p=\max\left(1,\sqrt{r_p m_i\nu_c
v_d/(Ze|V_p|)}\right)$. $F_E^p$ is always \emph{negative} or null,
because in the considered regime the downstream shadow of the dust
particle is fully ion-depleted~\cite{KhrapFluid1}. The ion impact part
of the ion-drag, depending on both the stream-lines topology and the
ion dynamics along those lines, can be decomposed in two parts. The
first corresponds to the flux of {\bf z}-momentum to the dust if the
ions were indeed cold~\cite{HutchRep}:
\begin{equation}
  F_{im,1}^p=\pi (r_p^2 n_{\infty} T_e) v_d
  \frac{|\phi_p|}{r_p\nu_c}\left(\tilde{a}_0^2+\frac{8}{3}+
  \frac{2}{\tilde{a}_0^2}-\frac{1}{3\tilde{a}_0^6}\right)\label{KhrapFi}
\end{equation}
The second accounts for the ion pressure at the dust surface, due to
the effective ion temperature along the stream-lines, \emph{never}
negligible even if $T_i\ll T_e$: $T_{i,\parallel}({\bf x})=T_i+m_i{\bf
v}({\bf x})^2$ (The heating depends on the local fluid velocity ${\bf
v}({\bf x})$ rather than ${\bf v_d}$, because the stream-lines are
curved in the dust vicinity). The ion temperature perpendicular to the
stream-lines is $T_{i,\perp}=T_i$. In the limit $T_i\ll T_e$,
integration of the ion pressure at the dust surface yields:
\begin{eqnarray}
  F_{im,2}^p=\frac{\pi}{6}(r_p^2 n_{\infty} T_e)\frac{m_i
  v_d^2}{T_e}\left[(6+8\alpha+3\alpha^2)\right.\nonumber\\
  -\tilde{a}_0^{-4}\left.(6-8\tilde{a}_0^{-2}\alpha+
  3\tilde{a}_0^{-4}\alpha^2)\right]
  \label{KhrapFi2},
\end{eqnarray}
where $\alpha=-ZeV_p/(m_i v_d\nu_c r_p)$ (we recall that $ZeV_p<0$);
$F_{im,1}^p$ and $F_{im,1}^p$ are
\emph{positive}. Eqs.~(\ref{KhrapCurr},\ref{KhrapFE},\ref{KhrapFi},\ref{KhrapFi2})
correspond to the mobility model.

\begin{figure}[!htp]
  \vspace{-0.4cm}
    \includegraphics[width=8.5cm]{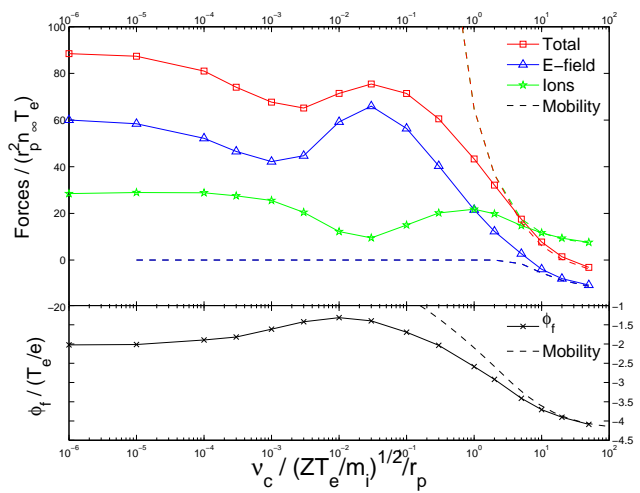}\\
    \vspace{-0.5cm}
    \caption{\small Ion-drag and dust floating potential against
      collisionality for $H^+$ ions; $T_i=0.01ZT_e$,
      $v_d=\sqrt{ZT_e/m_i}$ and $\lambda_{De}=20r_p$. The points
      correspond to SCEPTIC data at the dust surface, and the dashed
      lines to the mobility
      calculations~(\ref{KhrapFE},\ref{KhrapFi},\ref{KhrapFi2},\ref{KhrapCurr}).\label{KhrapComp2}}
    \vspace{-0.2cm}
\end{figure}
Fig.~(\ref{KhrapComp2}) is a plot of forces and floating potential
versus collisionality for $H^+$ ions with $T_i=0.01ZT_e$, that we now
analyze from high to low collisionality. In the continuum limit the
mobility model (dashed lines) agrees with SCEPTIC results (symbols),
thus validating both works. As the collisionality decreases, the
floating potential rises to a maximum at $\nu_c\simeq
0.01\sqrt{T_e/m_i}/r_p$ (a moderate collisionality tends to increase
the ion current~\cite{ScepticColl}). The E-field part of the ion-drag
follows a similar trend, i.e. peaks when $\lambda_{mfp}\lesim
\lambda_s$ (shielding length); the reason being that collisions favour
ion focusing downstream by breaking orbital angular momentum
conservation, hence increase the polarization field in the drift
direction. This effect, carefully explained in Ref.~\cite{Ivlev}, is
not taken into account by Maiorov~\cite{Maiorov} who therefore arrives
at the same erroneous conclusion as Schweigert (i.e. that a negative
ion-drag can occur at weak collisionality). The dip in ion-impact
force seen at $\nu_c\simeq 0.03\sqrt{T_e/m_i}/r_p$ is due to the local
floating potential maximum, whose effect is to reduce the energy at
which ions are collected. The local minimum in the E-field part of the
ion-drag at $\nu_c\sim 10^{-3}\sqrt{ZT_e/m_i}/r_p$ corresponds to
$\lambda_{mfp}\gesim \lambda_s$. As we keep reducing the
collisionality, the floating potential decreases to its
``collisionless'' ($\nu_{ii}\ll \nu_c \ll \sqrt{ZT_e/m_i}/r_p$) value,
causing the E-field part of the ion-drag to increase.

%%%%%%%%%%%%%%%%%%%%%%%%%%%%%
%%%% Khrapak Comp 1 %%%%%%%%%
%%%%%%%%%%%%%%%%%%%%%%%%%%%%%

Let us discuss further our results in a regime where the conditions
$b_{90}\ll\lambda_s$ and $v_d\ll v_{ti}$ are satisfied. This allows us
to directly compare our computations with the result of Ivlev and
coauthors, who calculate the electrostatic part of the ion-drag using
the linearized plasma equations, not accounting for downstream
depletion due to the finite-sized grain (appropriate at weak or
moderate collisionality)~\cite{Ivlev}:
\begin{equation}
  F_E^{p}(v_d)=(r_p^2 n_{\infty}
  T_e)\frac{ZT_e}{T_i}\phi_p^2\frac{4\sqrt{2}}{3}u
  \left[K\left(\frac{\lambda_D}{l_i}\right)
  +\sqrt{2\pi}\ln\Lambda\right]\label{KhrapFor1}
\end{equation}
Here $l_i=\sqrt{\pi/8}\lambda_{mfp}$,
$\ln\Lambda=\ln(\lambda_D/2b_{90})$, where $u=v_d/v_{ti}$. The
function $K$ is given by $K(x)=\atan(x)x+\left(\sqrt{\pi/2}-1\right)
x^2/(1+x^2)-\sqrt{\pi/2} \ln(1+x^2)$. We also used the vacuum
expression for the dust particle capacitance:
$V_p=Q_p/4\pi\epsilon_0r_p$.

\begin{figure}[!htp]
  \vspace{-0.4cm}
    \includegraphics[width=8.5cm]{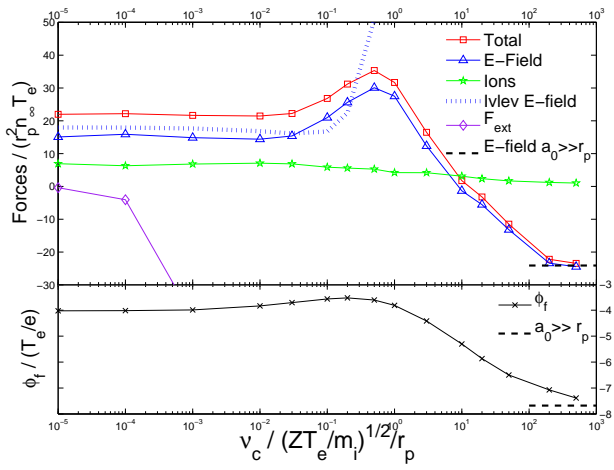}\\
    \vspace{-0.5cm}
    \caption{\small Ion-drag and floating potential versus
      collisionality for $Ar^{+}$ ions; $T_i=ZT_e$,
      $v_d=0.2\sqrt{ZT_e/m_i}$ and $\lambda_{De}=30r_p$. The points
      correspond to SCEPTIC data at the dust surface, the dotted line
      to Eq.~(\ref{KhrapFor1}), and the dashed lines to the
      limits~(\ref{KhrapCurrLim},\ref{KhrapFELim}). $F_{\ext}$ is the
      background electric force on the dust particle
      (Eq.~(\ref{ExtFor})), directed upstream.\label{KhrapComp1}}
    \vspace{-0.2cm}
\end{figure}

Fig.~(\ref{KhrapComp1}) is a plot of forces and floating potential
versus collisionality for equithermal $Ar^+$ ions and electrons,
$\lambda_{De}=30r_p$ and $v_d=0.2\sqrt{ZT_e/m_i}$. The ion-drag
components and dust floating potential follow a trend similar to what
has been observed in Fig.~(\ref{KhrapComp2}), although perhaps with a
more monotonic behaviour because at high temperature the ion flow is
less sensitive to slight variations in floating potential. Of course
the ion-drag peak is located at a higher collisionality
($\lambda_{mfp}\propto \sqrt{T_i}$ is larger here). The agreement
between SCEPTIC and Eq.~(\ref{KhrapFor1}) is only qualitative; recall
that Eq.~(\ref{KhrapFor1}) is only valid to logarithmic accuracy in
$\lambda_D/b_{90}$.

Fig.~(\ref{KhrapComp1}) also shows the high collisionality limit of
$\phi_f$ and $F_E$. While the mobility model does not apply for
$T_i=ZT_e$ (It requires $T_i\ll ZT_e$), it is valid in the limit
$\tilde{a}_0\gg 1$, i.e. when $E_{\ext}$ strongly dominates the dust
Coulombic field. In other words as $\tilde{a}_0\rightarrow\infty$:
\begin{eqnarray}
  I_i&\rightarrow&\pi n_{\infty} r_p^2
  v_d,\quad\textrm{and}\label{KhrapCurrLim}\\ F_E^p&\rightarrow& \pi
  r_p^2 n_{\infty} T_e \phi_p\label{KhrapFELim},
\end{eqnarray}
regardless of $T_i/ZT_e$ (provided of course the formula for vacuum
capacitance is applicable, i.e. $\lambda_{De}\gg r_p$). This proves
that past a certain collisionality, increasing $\nu_c$ does not
increase the negativity of $F_E$, bounded by:
\begin{equation}
  F_E^p\ge \pi (n_{\infty} r_p^2
  T_e)\ln\left(\frac{v_d\sqrt{\pi}}{2v_{te}}\right)\label{KhrapFELim2},
\end{equation}
where the logarithm is the floating potential calculated by equating
Eq.~(\ref{KhrapCurrLim}) with the collisionless electron current.

%For illustration purposes as well as for computational efficiency, the
%plasma parameters used in our figures have been choosen for the
%different force components to have the same magnitude. In typical
%dusty plasmas however $\lambda_s\gg r_p$, hence exept in the continuum
%limit the main contribution to the ion-drag comes from $F_E^p$
%(E-field part of the ion-drag).

We have therefore shown that only in the continuum limit
($\nu_c\gg\sqrt{ZT_e/m_i}/r_p$) can the ion-drag reverse. There the
ambipolar electric force on the dust particle (${\bf
F_{\textrm{ext}}}=Q_p{\bf E_{\textrm{ext}}}$)
\begin{equation}
  {\bf F_{\ext}} \sim 4\pi\phi_p(r_p^2 n_{\infty} T_e)
  \frac{\lambda_{De}^2}{r_p^2}
  \frac{\nu_c}{\sqrt{ZT_e/m_i}/r_p}\frac{v_d}{\sqrt{ZT_e/m_i}}\label{ExtFor}
\end{equation}
is much larger than ion-drag itself (Fig.~(\ref{KhrapComp1})); in the
typical situation where $Q_p\sim 4\pi\epsilon_0r_p V_p<0$, ${\bf
F_{\textrm{ext}}}$ is directed upstream.

We have also performed extensive simulations, not shown here, where
the ion drift is driven by a neutral flow assumed unaffected by the
dust particle (kinetic regime); the ion distribution at infinity is
then given by a drifting Maxwellian rather than by
Eq.~(\ref{fionSn}). The ion-drag is very similar in magnitude to the
results presented here, for comparable collisionality and flow
velocity. The negative $F_{\textrm{ext}}$ (Eq.~(\ref{ExtFor})) is
however replaced by a positive neutral drag.

% whose magnitude when $\nu_c\gg \sqrt{ZT_e/m_i}/r_p$ is much larger
%than the ion-drag unless . So in typical flowing-neutral cases, the
%total force on the particle is never in the upstream direction because
%ion-drag reversal occurs only when the other downstream forces are
%dominant.

%In a typical atmospheric pressure argon plasma however
%($T_i=T_n=300K$, Neutral pressure: $P=10^5Pa$, Ion/Electron density:
%$n_{\infty}=10^{18}m^{-3}$, gas viscosity: $\eta\sim 2\cdot
%10^{-5}Pa\cdot s$, CX cross section: $\sigma_c\sim 10^{-15}cm^2$, Mass
%$m_i=m_n=40m_{H}$), the neutral flow around a micron-size dust
%particle is in the viscosity regime ($r_p=1\mu m\gg
%\lambda_{mfp,n}$). Therefore even when $\tilde{a}_0\gg 1$ the neutral flow
%populates the downstream shadow of the dust particle, and $|F_E^p|$ is
%much smaller than predicted by Eq.~(\ref{KhrapFELim2}). Let us however
%disregard this inconsistency and consider the neutral drag to be given
%by the hydrodynamic Stokes formula:
%\begin{equation}
%  |{\bf F_{Stokes}}| \sim 6\pi (r_p^2 n_{\infty} T_e)
%   \frac{\eta\sqrt{ZT_e/m_i}}{r_p n_{\infty}
%   T_e}\bar{v}_d\label{Stokes}.
%\end{equation}
%For the considered parameters $\eta\sqrt{ZT_e/m_i}/r_p n_{\infty} T_e\sim 10^6$. In other
%words we need the drift velocity to be $10^6$ times smaller than the
%sound speed for the neutral drag to be of comparable magnitude to the
%E-field part of the ion-drag force!
%  

In conclusion, we have presented fully self-consistent calculations of
the ion-drag force on a spherical dust particle over the entire range
of charge-exchange collisionality, when the ion flow is driven by an
external electric field. We have shown that the ion-drag behaves
similarly to the floating potential; it shows a local maximum when
$\lambda_{mfp}\lesim \lambda_s$, and tends to a limit at high
$\nu_c$. Although there are physically meaningful parameters where the
ion-drag is negative, that occurs only in collisional regimes where
the direct electric field force far exceeds the drag.

%%%%%%%%%%%%%%%%%%%%%%%%%%%%%
%%%% Acknowledgments %%%%%%%%
%%%%%%%%%%%%%%%%%%%%%%%%%%%%%

\begin{acknowledgments}

  Leonardo Patacchini was supported in part by NSF/DOE Grant
  No. DE-FG02-06ER54891.
%The SCEPTIC calculations were performed on
%  the Alcator Beowulf cluster which is supported by U.S. DOE Grant
%  No. DE-FC02-99ER54512.

\end{acknowledgments}

\end{document}